\documentclass[twocolumn,prb,showpacs,multicol,amsmath,amssymb]{revtex4}
\usepackage[dvips]{graphicx}
\usepackage{graphicx}
\usepackage{dcolumn}
\usepackage{bm}
\usepackage{graphics}
\usepackage{epsfig,color}

\newcommand{\be}{\begin{equation}}
\newcommand{\ee}{\end{equation}}
\newcommand{\bea}{\begin{eqnarray}}
\newcommand{\eea}{\end{eqnarray}}

\begin{document}
\title{From fidelity to entanglement of entropy of the one-dimensional transverse-field quantum compass model}
\author{Mostafa Motamedifar, Saeed Mahdavifar, Somayyeh Nemati, Saber Farjami Shayesteh}
\affiliation{ Department of Physics, University of
Guilan,41335-1914, Rasht, Iran}
\date{\today}
\begin{abstract}
We study fidelity and fidelity susceptibility by addition of entanglement of entropy in the one-dimensional quantum compass model in a transverse magnetic field. All four recognized gapped regions in the ground-state phase diagram (GSPD) are in the range of our calculation. We show that the difference between the position of the sharp drop of fidelity $h^{*}$ from real critical field $h_{c}$ is inversely related to the number of particles ($N$) with a relation such as $h^{*}=h_{c}+\Sigma_{j} c_{j} N^{-1/\nu}$ for a finite chain. In this relation $\nu$ is the correlation length critical exponent. The scaling behavior of the extremum of fidelity susceptibility shows that the amount of $\nu$ depends on the selected area of GSPD. Furthermore, we calculated a recently proposed quantum information theoretic measure, Von-Neumann entropy, and show that this measure provides appropriate signatures of the quantum phase transitions (QPT)s occurring at the critical fields. We calculated Von-Neumann entropy between one-, two- and three-particle blocks with the rest of the system. We show that in an alternating model such as quantum compass model, the value of entanglement between a two-particle block with the rest of the system is more dependent on the power of exchange couplings connecting the block with the rest of the system than the power of exchange coupling between two particles in the own block. In other words, a pair with a strong coupling does not see the rest of the system. Also, in some areas of GSPD, amount of entanglement of a two-particle block in an odd link is the same as that of an even link in a factorized point independent on where the block is.
\end{abstract}
\maketitle
\section{Introduction}\label{sec1}
Recently, much attention has been paid to the role of orbital degrees of freedom in some materials such as transition metal compounds\cite{cheong07, tokura00, wakabayashi06}. In some these materials, the \emph{d} orbital degeneracy of transition metal ions is incompletely lifted, and the remaining orbital degrees of freedom can be generally characterized by localized S=1/2 pseudo spins, which have been bond-selective interactions\cite{Wen-Long2011}. Subsequently; a frustration induces in the interactions among different bonds\cite{Wen-Long2011,mishra2004}. Mott insulators \cite{kugel73,kim2008,jackeli2009} such as Sr$_{2}$IrO$_{4}$, Colossal magnetoresistances\cite{khomski2003}, high-temperature superconductivities\cite{nussinov05} and so on indicate many significant roles of orbital degrees of freedom.

A proposal model for describing quantitatively the nature of the orbital states is the \emph{quantum compass model}. The observed microscopic treatment of Mott-insulators with orbit degeneracy by a pseudo-spin caused that this model be suggested\cite{kugel73}. However the quantum compass model has been developed to any materials with complex interplay between orbital, spin, charge and lattice degrees of freedom. In the quantum compass model, the orbital degrees of freedom are represented by pseudo-spin operators and coupled anisotropically in such a way as to mimic the competition between orbital ordering in different directions, where the coupling along one of the bonds is Ising type, but different spin components are active in other bonds directions\cite{erik09,Breziskii2007}.
\begin{figure}[t]
\centerline{\psfig{file=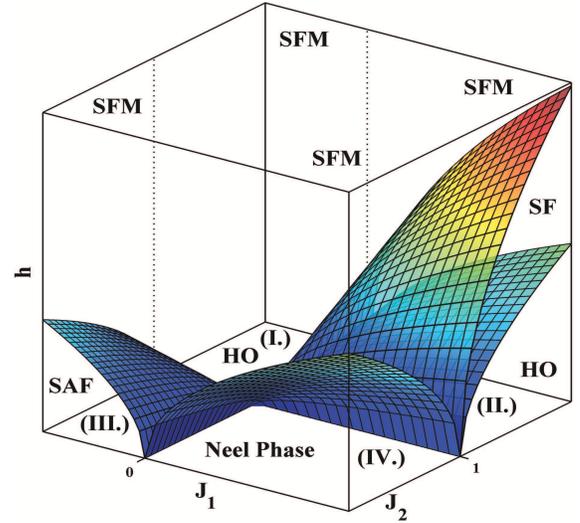,width=7.5cm,height=7.0cm}}
\caption{(Color online.) the GSPD of the QCC in the absence a magnetic field and the presence of it\cite{motamedi11}}. \label{phase}
\end{figure}

The one-dimensional quantum compass model, known as the \emph{quantum compass chain} (QCC), is exactly solvable by mapping to quantum Ising model\cite{Breziskii2007}. Its GSPD is extremely rich as it exhibits interesting properties while approaching to the quantum critical point at zero temperature\cite{erik09} and induces various fascinating physical phenomena. Characteristic features of interplay spin-orbital in the QCCs can play a dominant role in the quantum information theory and enhanced quantum effects and may lead to entangled spin-orbital ground state\cite{oles06}. In the absence of the magnetic field, the GSPD is divided to four gapped regions\cite{erik09, mahdavifar10}. These regions are separated by two interesting transition lines. A line of the first-order phase transition crosses with a line of the second-order phase transition. The point of joint of them is known the \emph{multicritical point}.

QPTs take place when controlling parameter changes across critical point, and some properties of the many-body system will change dramatically\cite{sachdev11}. Therefore, since the exotic magnetic behaviors at the QPTs are observed, many scientists are interested to research phenomena that appear QPTs. During the past few years, some important concepts in quantum information theory have been introduced to characterize QPTs\cite{osterloh02}. For instance, entanglement, which is one of the main concepts in quantum information theory, can offer a useful signature for some QPTs\cite{osterloh02}. Several measures of entanglement have been used to investigate states of matter\cite{osterloh02, vidal03}. Among them, the Von-Neumann entanglement entropy (EE) quantifies the bipartite entanglement between two parts of a quantum system\cite{bennett96}. Besides entanglement; fidelity is another quantum information concept, which has also been applied in characterizing QPTs. Fidelity is an overlap between two quantum states and can measure the similarity between them\cite{wootters81, schumacher95, josa94}.
\begin{figure*}[t]
\centerline{\includegraphics[width=17cm,height=14cm,angle=0]{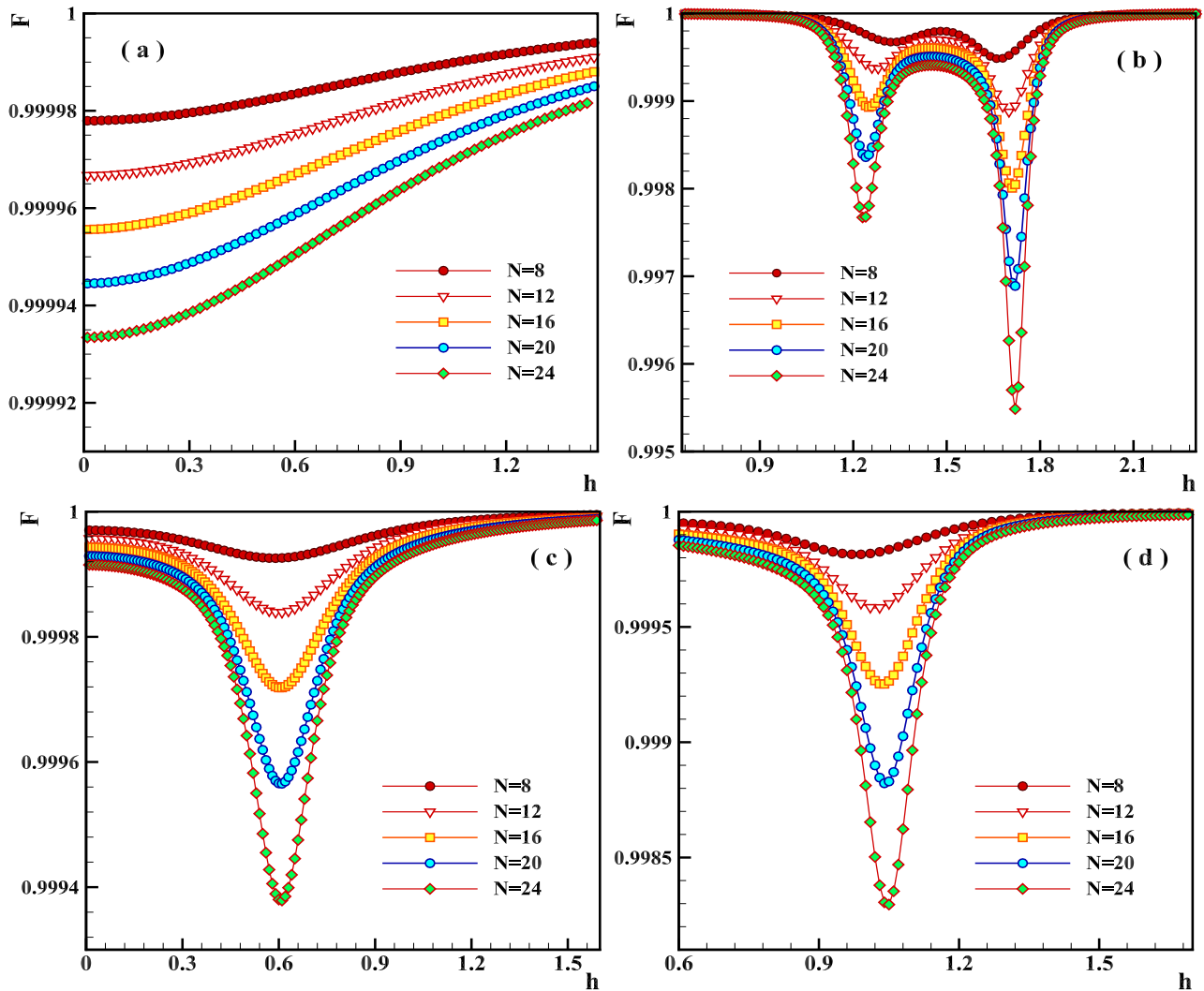}
}\caption{(Color online.) The Fidelity as a function of the transverse magnetic filed $h$, for different chain lengths $N=8, 12, 16, 20, 24$ and exchanges $L_{1}=1.0$, (a) $J_1=-3.0,J_2=3.0$, (b) $J_{1}=3.0, J_2=3.0$, (c) $J_{1}=-3.0, J_2=0.5$ and (d)
$J_{1}=3.0, J_2=0.5$ .} \label{fidelity-complete}
\end{figure*}

In this work, it would be interesting to investigate the fidelity and fidelity susceptibility on the ground state of QCC for the all of the GSPD of QCC in the presence of a magnetic field in the framework of Lanczos method. The first point worth to emphasize is, Ke-Wei Sun and Quing-Hu Chen\cite{ke-wei2009} have investigated fidelity and fidelity susceptibility of QCC, but  their results solely restricted to move on the first and the second-order phase transition lines\cite{erik09, mahdavifar10} of the GSPD of QCC. Their studies couldn't cover the all of the GSPD of QCC. The second point worth to emphasize is that our results obtained from the finite size scaling techniques are a bit different from the achieved critical exponents in Ref.[25]. Probably, it seems that the some corrections are needed for its results about the correlation length critical exponent. In addition, in this paper, we have noted to entanglement of entropy of QCC in a new point of view by numerical Lanczos method.

This paper is organized as follows. In section II, besides introducing the Hamiltonian of the one-dimensional quantum compass model, we will assign to the type of phases of QCC which recognized\cite{motamedi11} in the presence of a magnetic field. In section III, we present our numerical results, which include the fidelity and fidelity susceptibility in different regions of the GSPD. In addition, Von-Neumann entropy is noted in section IV.
Finally, a summary is presented in section V.

\section{ Numerical method} \label{sec2}
\begin{figure*}[t]
\centerline{\includegraphics[width=18cm,height=8.0cm,angle=0]{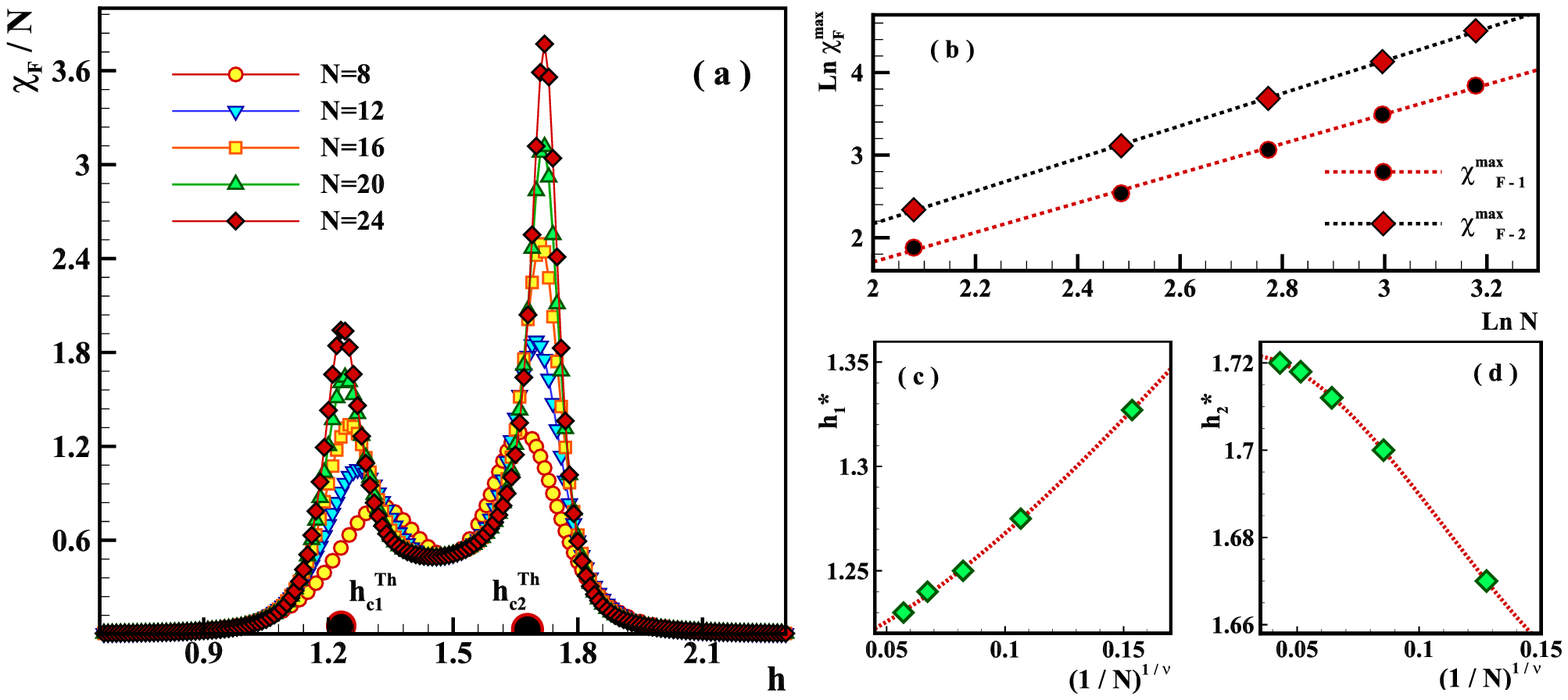}
}\caption{(Color online.) (a) The fidelity susceptibility as a function of the transverse magnetic field $h$ for chain length $N=8, 12, 16, 20, 24$ and exchanges $L_{1}=1.0$, $J_1=3.0, J_2=3.0$ for the 2nd region of the GSPD ($h^{Th}_{c1}$ and $h^{Th}_{c2}$ are the theoretical results for critical fields obtained from exact solution\cite{motamedi11}), (b) $\ln\chi_{F}^{max}$ versus $\ln N$ for both of its critical points. (c) and (d) the best polynomial fitting of $h^{*}$ versus $1/N$ respectively for the first critical point and the second one.
} \label{FS-II}
\end{figure*}

Our numerical experiments have been focused on the ground state phases of the QCC by considering a uniform magnetic field as a control parameter.
The Hamiltonian of this model in the presence of a transverse magnetic field with periodic boundary condition is given by\cite{jafari11, motamedi11}:
\begin{eqnarray}
{\cal H} &=& \sum_{i=1}^{N/2} J_{1}S_{2i-1}^{z}S_{2i}^{z}+ J_{2}S_{2i-1}^{x}S_{2i}^{x}+ L_{1}S_{2i}^{z}S_{2i+1}^{z}\nonumber\\
&-& h  \sum_{i=1}^{N}S_{i}^{y}\ , \label{Hamiltonian}
\end{eqnarray}
where $S=\frac{1}{2}\sigma_{i}^{\alpha}$ and $\sigma_{i}^{\alpha}$ is the Pauli matrix in direction $\alpha(x,y,z)$ on site $i$, $N$ is the number of the sites and $J_{1}, J_{2}, L_{1}$ are the exchange couplings.
One of the most accurate numerical methods for the study of zero temperature behavior is the Lonczos method which is used to diagonalize exactly quantum compass chains with lengths up to $N=24$ and periodic boundary conditions for different values of the exchanges. It has been demonstrated that in the absence of any magnetic field, there are four regions based on exchange measures\cite{erik09,mahdavifar10}: $(I.) J_{1}/L_{1}<0, J_{2}/L_{1}>1$ (1st), $(II.) J_{1}/L_{1}>0, J_{2}/L_{1}>1$ (2nd), $(III.) J_{1}/L_{1}<0, J_{2}/L_{1}<1$ (3rd), $(IV.) J_{1}/L_{1}>0, J_{2}/L_{1}<1$ (4th).
Picking out various exchange couplings, the system may be placed in one of the various regions of the GSPD. As Fig.~\ref{phase} presents, in the presence of a magnetic field the system in the region (I.) shows a phase with a hidden order (HO) and a saturated ferromagnetic (SFM) phase and there are no quantum phase transitions; furthermore, the system in the region (II.) exhibits three quantum phases with different order parameters in the presence of a magnetic field. HO phase is followed by a spin-flop (SF) phase, and the final phase is SFM. So, two QPTs have been recognized in the region (II.).  Each of the regions (III.) and (IV.) has one critical point connecting respectively stripe antiferromagnetic (SAF) phase and N\'{e}el phase to a saturated ferromagnetic phase\cite{motamedi11}.
\begin{figure*}[t]
\centerline{\includegraphics[width=18.0cm,height=11.5cm,angle=0]{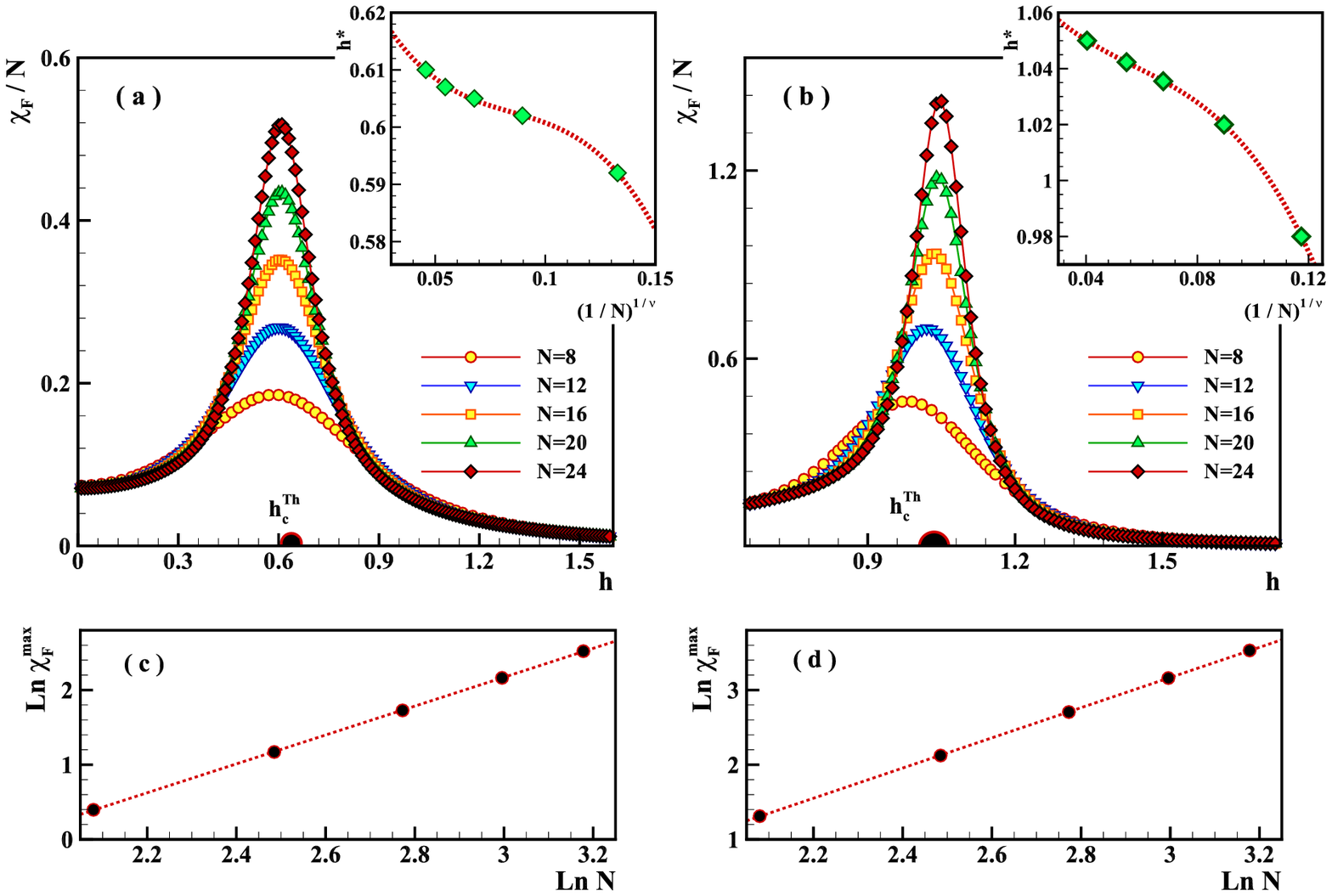}
}\caption{(Color online.) The fidelity susceptibility as a function of the transverse magnetic field $h$ for chain length $N=8, 12, 16, 20, 24$, and their insets indicating the best polynomial fitting of $h^{*}$s versus $1/N$ for exchanges (a)$J_1=-3.0,J_2=0.5$ related to the 3rd region of the GSPD,(b)$J_{1}=3.0, J_2=0.5$ regard to the 4th region of the GSPD (c) $\ln\chi_{F}^{max}$ versus $\ln N$ for the 3rd region and (d) $\ln\chi_{F}^{max}$ versus $\ln N$ for the 4th region.} \label{IIIvIV}
\end{figure*}
\section{FIDELITY and Fidelity susceptibility}
\subsection{Fidelity}
 Here, we are going to look at the fidelity function and try to establish a bridge between QPTs and fidelity in considerable detail through the QCC. As we know, a QPT identifies any point of non-analyticity in the ground state energy of an infinite lattice system \cite{sachdev11}. Conventionally, local order parameters are needed to detect the non-analyticity in the ground state properties as the system varies across the quantum critical point (QCP). However, the knowledge of the local order parameter is not easy to retrieve from a general many-body system for which Eq.~(\ref{H-fidel}) can in general be written as Hamiltonian:
 \begin{eqnarray}
{\cal H}= {\cal H}_{0}+\lambda {\cal H}_{\lambda} ,
 \label{H-fidel}
\end{eqnarray}
where $\lambda$ is a variable which typically parameterizes an interaction and exhibits a phase transition at some critical value $\lambda_{c}$. In this form ${\cal H}_{\lambda}$ is then recognized as a term that drives the phase transition\cite{w-L2007}.
 Recently, quantum fidelity, also referred to as the ground state fidelity, sparked great interest among the community to use it as a probe for the QCP \cite{quan06,zanardi06}.
 The fidelity is given by the modulus of the overlap of normalized ground state wave functions $|\psi(\lambda)\rangle$ and $|\psi(\lambda+\delta\lambda)\rangle$ for closely spaced Hamiltonian parameter $\lambda$ and $\lambda+\delta\lambda$. In other words, the fidelity defines the overlap between two neighboring ground states of a quantum Hamiltonian in the parameter space\cite{gu2010}, i.e.,
\begin{eqnarray}
F(\lambda,\lambda+\delta\lambda)=|\langle\psi(\lambda)|\psi(\lambda+\delta\lambda)\rangle| . \label{fidelity}
\end{eqnarray}
Here we have selected the magnetic field as a space Hamiltonian parameter $\lambda$. Fig.~\ref{fidelity-complete} shows the ground state fidelity of the QCC as a function of $h$ with parameter interval $\delta h=0.01$ for different regions of the GSPD. The presented numerical results in Fig.~\ref{fidelity-complete} cover all four regions and regard chains with lengths $N=8, 12, 16, 20, 24$. As shown in Fig.~\ref{fidelity-complete} for the first region of the GSPD in which no QPTs occur, there is a size effect with the coefficients of the ground state eigenvector in the absence of a magnetic field. Increasing the magnetic field vanishes this size effect in a way that overlapping of two deferent neighbor ground states will enhance as the value of fidelity close to one in large enough magnetic fields. In contrast, the results of fidelity's calculation of the region (II.) shown in Fig.~\ref{fidelity-complete}(b) indicate that in this region of the GSPD and in small magnetic fields ($h\rightarrow 0$) there is no size effect on the coefficient of the ground state eigenvector. It seems that the sharp drops of fidelity can be described with a dramatic change in the structure of the ground state of the system during the QPTs. Away from these points, the fidelity almost equals to unity, and in other words, the ground states overlap to each other completely.
In Fig.~\ref{fidelity-complete}(c) and Fig.~\ref{fidelity-complete}(d) we have plotted the fidelity function related to regions (III.) and (IV.) respectively. In each of these figures, one can obviously recognize an abrupt drop for the 3rd and 4th regions of the GSPD correspondingly.
As shown in three last figures, the number of particles greater, the drops at the critical fields sharper. In fact, at the critical points since the ground state eigenvector replaced with another orthogonal quantum state eigenvector, the fidelity between two different ground states will near to zero in the thermodynamic limit no matter how small the difference in parameter $\delta h$ is. Therefore, it is concluded that the fidelity can describe quantum phase transition in its own way. However, how close to the really critical fields from the finite size scaling is assigned to the next section.
\subsection{Fidelity Susceptibility}
Though borrowed from the quantum information theory, fidelity has been proven to be a useful and powerful tool to detect and characterize QPTs in condensed matter physics\cite{gu2010}. In order to remove the artificial variation of external parameters, the concept of fidelity susceptibility is introduced\cite{w-L2007}.
Using Eq.~(\ref{fidelity}), a series expansion of the ground state fidelity, can then be written as:
\begin{eqnarray}
 F(\lambda)= 1-\frac{(\delta \lambda)^2}{2}\frac{\partial^{2}F}{\partial {\lambda}^2} ,
 \label{H-fidel-dif}
\end{eqnarray}
where $\partial^{2}_{\lambda}F\equiv\chi_{F}$ is called the fidelity susceptibility. If the higher-order terms are taken to be negligibly small, then the fidelity susceptibility is defined as
\begin{eqnarray}
\chi_{F}&=&\frac{2(1-F(\lambda))}{\delta\lambda^{2}}
\equiv \lim_{ \delta\lambda\rightarrow0}\frac{-2\ln F}{(\delta\lambda)^{2}} .
 \label{sf-dif}
\end{eqnarray}

Figures \ref{FS-II} (a), \ref{IIIvIV} (a) and \ref{IIIvIV}(b) show the numerical results of the per site fidelity susceptibility of the QCC in the 2nd, 3rd and 4th region of the GSPD respectively. It can be seen that the averaged fidelity susceptibility for different $N$ all show peaks at pseudo-critical fields $h^{*}$ in where $\chi_{F}/N$ becomes more pronounced for more particles ($N$). We can deduce that at these pseudo-critical points $\chi_{F}/N$ is an extensive quantity. However, it can be seen that on both sides around the these points (off-critical fields), the averaged fidelity susceptibility is an intensive quantity, i.e. $\chi_{F}\sim N$.
\begin{table}[b]%
\caption{Correlation length exponents for different regions of the GSPD.}
\label{tablenu}\centering%
\begin{tabular}{llc}
\hline\hline %
\multicolumn {2}{c}{region of the GSPD} &  correlation length exponent\\
       &                                & $(\nu)$\\[1ex]\hline%
II     & the 1st critical point         & $1.11\pm0.01$\\\cline{2-3}
       & the 2nd critical point         & $1.01\pm0.01$\\[1ex]\hline
III    &                                & $1.03\pm0.01$\\[1ex]\hline
IV     &                                & $0.99\pm0.01$\\[1ex]\hline\hline
\end{tabular}
\end{table}
From the scaling relation\cite{zanardi2007,capponi2009,capponi2010}$\chi_{F(h^{*})}\sim N^{2/d\nu}$ at the pseudo-critical point, a linear dependence for the maximum fidelity susceptibilities $\chi^{max}_{F}$ on $N^{2/d\nu}$ is expected. The critical exponent $\nu$, is the correlation length exponent at the critical point and $d$ is the dimension of the system that here $d=1$. This is confirmed by the results shown in Fig. \ref{FS-II}(b), in which $\ln \chi^{max}_{F}$ scales linearly with $\ln N$ at pseudo-critical points, and the slope is related to the critical exponent of the correlation length as $2/\nu$. The results related to calculation of correlation length exponents for various region of the GSPD are shown in the Table.\ref{tablenu}.

On the other hand, Zanardi and \emph{et.al}\cite{zanardiA2008} claimed that the difference between the position of the pseudo-critical point ($h^{*}$) and the real critical point ($h_{c}$) is inversely related to the number for a finite chain with a relation such as $|h^{*}-h_{c}|\sim N^{-1/\nu}$. With the increasing of the size $N$ the positions of pseudo-critical point ($h^{*}$) come close to the really critical points. Where $L=N^{1/d}$ is the system size which is related to the number of particles $N$, and $d$ standing for dimensionality. Our calculations show that this relation needs some more corrections.
As shown in the Figs. \ref{FS-II}(c) and (d) related to the 2nd region of the GSPD of QCC, the peak positions of pseudo critical fields $h^{*}$s are plotted as a function of $(1/N)^{1/\nu}$. Also, the insets of the Figs. \ref{IIIvIV}(a) and (b) are devoted to the best polynomial fitting of data for the 3rd and 4th region of the GSPD. These best  polynomial fittings are listed in the Table.\ref{tablefitting}.
\begin{table}[b]%
\caption{The functions of the pseudo critical fields versus $N^{1/\nu}$ for the different regions of the GSPD.}
\label{tablefitting}\centering%
\begin{tabular}{llc}
\hline\hline %
\multicolumn {2}{c}{region of the GSPD} & position of pseudo critical fields \\
       &                                & $(h^{*})$\\[1ex]\hline%
II     & the 1st critical point         & $h^{*}_{1}\approx1.202+0.241 N^{-1/\nu_{c1}}$\\
       &                                & $+5.029 N^{-2/\nu_{c1}}-8.473 N^{-3/\nu_{c1}}$\\[1ex]\cline{2-3}
       & the 2nd critical point         & $h^{*}_{2}\approx1.714-0.587 N^{-1/\nu_{c2}}$\\
       &                                & $-11.827 N^{-2/\nu_{c2}}+35.482 N^{-3/\nu_{c2}}$\\[1ex]\hline
III    &                                & $h^{*}\approx0.641-1.151 N^{-1/\nu}$\\
       &                                & $+12.245 N^{-2/\nu}-48.063 N^{-3/\nu}$\\[1ex]\hline
IV     &                                & $h^{*}\approx1.096-1.888 N^{-1/\nu}$\\
       &                                & $-24.322 N^{-2/\nu}-141.820 N^{-3/\nu}$\\[1ex]\hline\hline
\end{tabular}
\end{table}
\begin{figure}
\centerline{\psfig{file=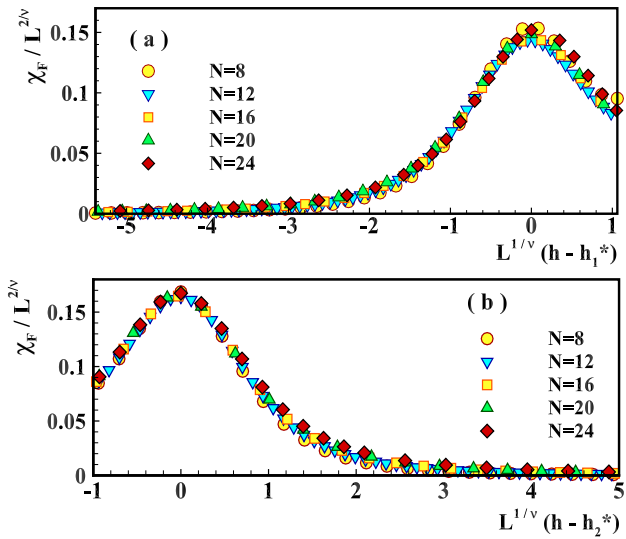,width=9.5cm,height=8.5cm}}
\caption{(Color online.) Data collapse for the $\chi_{F}/L^{2/\nu}$ for different chain lengths
    $N=8, 12, 16, 20, 24$ and exchanges $L_{1}=1.0$, $J_1=3.0, J_2=3.0$, 2nd region of the GSPD (a) for the first critical point $\nu_{c1}=1.11$. (b) for the second critical point $\nu_{c2}=1.01$} \label{1scaling}
\end{figure}

This analysis shows that the $|h^{*}-h_{c}|\sim N^{-1/d\nu}$ should be replaced with $|h^{*}-h_{c}|\approx c_{1}N^{-1/d\nu}+c_{2}N^{-2/d\nu}+...$. Where $c_{1}$ and $c_{2}$ are the coefficients.
Comparing this corrected relation with our results we can obtain $h_{c1} = 1.202,  h_{c2} = 1.714$ for the 2nd region of the GSPD, and $h_{c} = 0.641$ and $h_{c} = 1.096$, for the 3rd and 4th region of the GSPD in the thermodynamic limit. Comparing these values of critical points to those ($h^{Th}_{c}$)s obtained from us previous work\cite{motamedi11} ($\frac{\sqrt{J_{1}(L+J_{2})}}{2}$ and $\frac{\sqrt{-J_{1}(L-J_{2})}}{2}$), our results, here are well consistent with the exact critical points.
\begin{figure}[t]
\centerline{\psfig{file=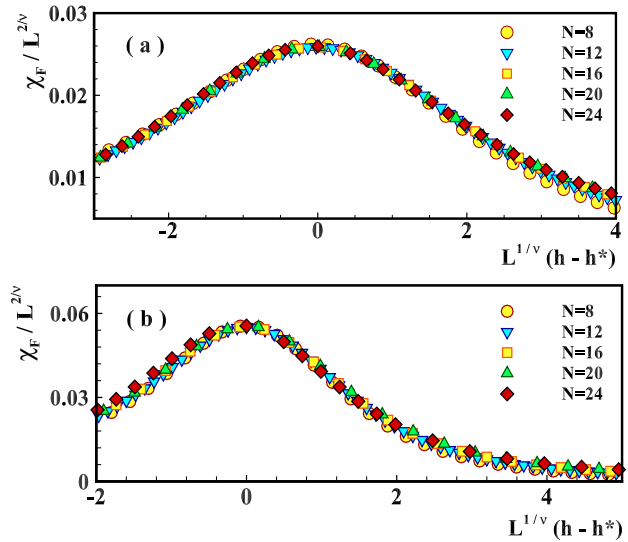,width=9.5cm,height=8.5cm}}
\caption{(Color online.)Data collapse for the $\chi_{F}/L^{2/\nu}$ for different chain lengths $N=8, 12, 16, 20, 24$ and (a) exchanges $L_{1}=1.0$, $J_1=3.0, J_2=-0.5$, the 3rd region of the GSPD  $\nu_{c}=1.03$ and (b) exchanges $L_{1}=1.0$, $J_1=3.0, J_2=0.5$, the 4th region of the GSPD  $\nu_{c}=0.99$ } \label{2scaling}
\end{figure}
In the checking the correctness of obtained $\nu$ one way is investigating the finite-size scaling behavior of fidelity susceptibility. For this reason, we are following a scaling technique in which all graphs collapse on each other. The scaling technique is based on the divergence of fidelity susceptibility closed to the critical points (Figs. \ref{FS-II}) and the power law scaling from previous discussion, we expect that the behavior of $\chi_{F}$ on finite systems in the neighborhood of quantum critical point be defined\cite{capponi2010,sandvik2011} as $\chi_{F}=L^{2/\nu} f_{\chi_{F}}(L^{1/\nu}|h-h^{*}|)$ where $f$ is an unknown regular scaling function. The obtained data collapse plots are displayed in Fig .\ref{1scaling} in which  (a) the used value of $\nu$ is $1.11$ and  (b) the value of $\nu$ is $1.01$.
On the other hand, to check the scaling function and correctness of previous results of the 3rd and 4th region, the obtained data collapse plots are displayed in Fig .\ref{2scaling} in which (a) the used value of $\nu$ is $1.03$ and (b) the value of $\nu$ is $0.99$.
\section{Entanglement of Entropy}
\begin{figure}[t]
\centerline{\psfig{file=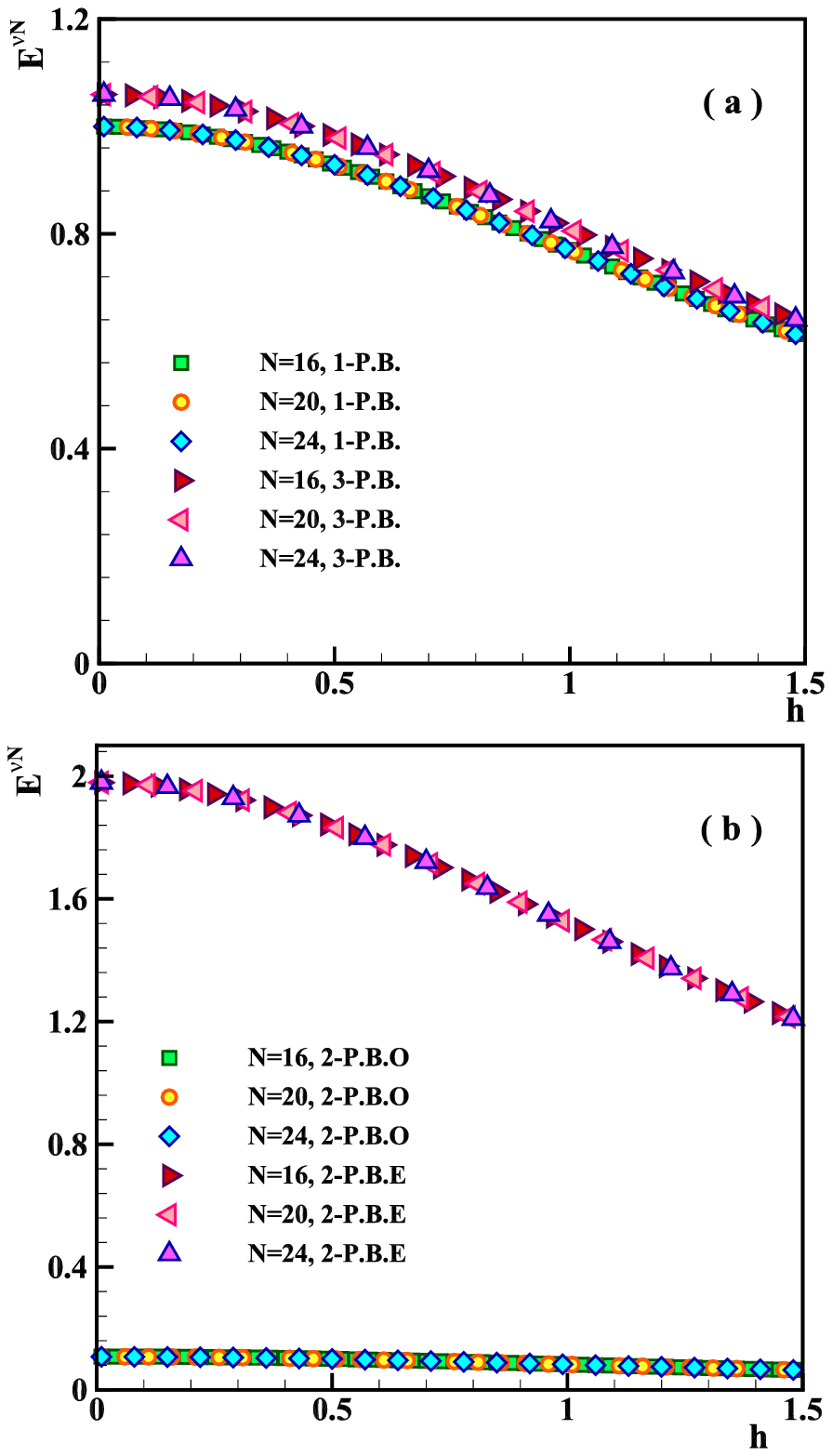,width=8.0cm,height=15.0cm}}
\caption{(Color online.)Entanglement of entropy as a function of the transverse field $h$ for different chain lengths
    $N=16, 20, 24$ and exchanges $L_{1}=1.0$, $J_1=-3.0, J_2=3.0$ (a) entanglement of one-particle block (1-P.B.) and three-particle block (3-P.B.) with the rest of system. (b) entanglement of two-particle blocks located in odd (2-P.B.O) and even links (2-P.B.E) with the rest of system} \label{EE-Na-4}
\end{figure}
Entanglement of entropy quantifies the entanglement between a block of $L$ contiguous spins and the rest of the chain (ROC) and defined as:
\begin{eqnarray}
E^{\nu N} &=-\langle \mathrm{log} ~\hat{\rho}_{A}\rangle =-Tr_{A}[\hat{\rho}_{A}~\mathrm{log}~\hat{\rho}_{A}] , \label{EE}
\end{eqnarray}
where ~$\hat{\rho}_{A} = Tr_{B}[\hat{\rho}_{A}]$, and $\hat{\rho}$ is the density matrix of the ground state. It is assumed that the system consists of subsystems A and B. The entanglement of entropy (EE) quantifies the information describing the entanglement between the subsystems A and B.
For QCC in which a unit cell consists of three particles, entanglement between various multi-spin blocks and the ROC can be studied. One-particle, two-particle and three-particle blocks are properly selected for a numerical experiment. In addition, because of the existence of different exchange couplings in odd and even links, it is expected that various two-particle blocks in different links have different EE behaviors. In this section, we set to calculate the EE in all GSPD of QCC and for various blocks, however, to prevent of redundancy, the scaling behavior of this quantity is assigned to another work\cite{motamedi12}.

For the 1st region of the GSPD, Figs.~\ref{EE-Na-4}(a) and (b) are devoted to describe the behavior of EE between selected blocks and ROC. The Fig.~\ref{EE-Na-4}(a) describes EE treatment between one-particle and three-particle blocks with ROC. This plot presents that in the absence of a magnetic field both selected blocks are entangled with the ROC. Despite the different amounts, EE has a similar descending behavior for different blocks.

Furthermore, Fig.~\ref{EE-Na-4}(b) is devoted to EE behavior of a two-particle block placed on an odd and an even link respectively. This figure presents that a two-particle block located at an odd link is entangled with the ROC hundred times as much as a two-particle block of an even link does.
\begin{figure}[t]
\centerline{\psfig{file=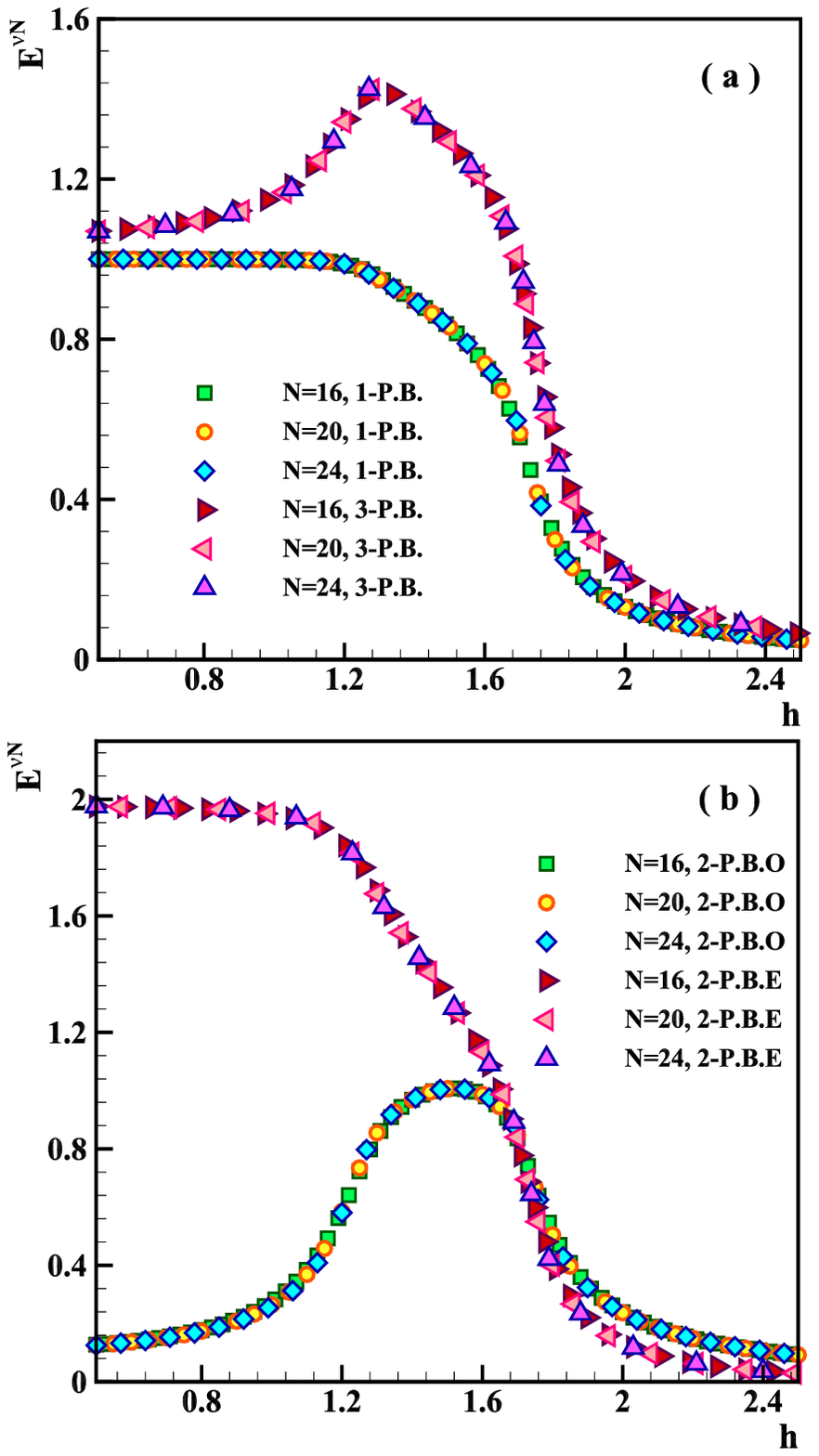,width=8.0cm,height=15.0cm}}
\caption{(Color online.)Entanglement of entropy as a function of the transverse field $h$ for different chain lengths
    $N=16,20,24$ and exchanges $L_{1}=1.0$, $J_1=3.0, J_2=3.0$ (a) entanglement of one-particle block (1-P.B.) and three-particle block (3-P.B.) with the rest of system. (b) entanglement of two-particle blocks located in odd (2-P.B.O) and even links (2-P.B.E) with the rest of system.} \label{EE-Na-1}
\end{figure}
In other words, a pair with a strong coupling does not see the rest of the system. It indicates the value of entanglement between a two-particle block with the rest of the system is more dependent on the power of exchange couplings connecting the block with the ROC than the power of exchange coupling between two particles in the own block. Odd link couplings ($J_{1},J_{2}$) are stronger, but the calculated EE value of them is very smaller than those of even links ($L_{1}$) having relatively weaker couplings.
By increasing a magnetic field, never again does the entanglement of entropy reach to the values related to the absence of a field. It means that for this region of the GSPD, every block shows a reduction treatment in the presence of a magnetic field.
\begin{figure}[t]
\centerline{\psfig{file=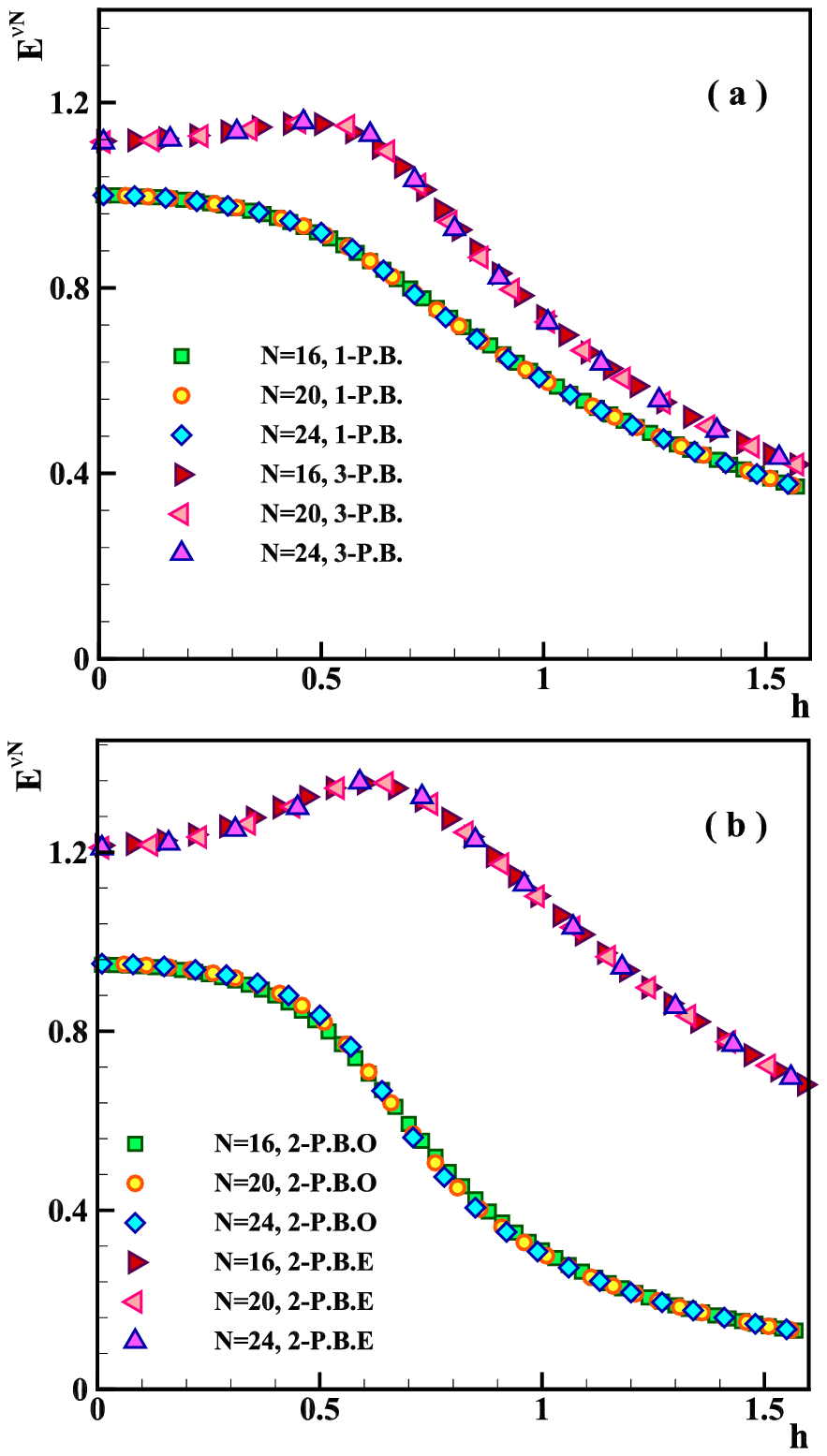,width=8.0cm,height=15.0cm}}
\caption{(Color online.)Entanglement of entropy as a function of the transverse field $h$ for different chain lengths
    $N=16,20,24$ and exchanges $L_{1}=1.0$, $J_1=-3.0, J_2=0.5$ (a) entanglement of one-particle block (1-P.B.) and three-particle block (3-P.B.) with the rest of system. (b) entanglement of two-particle blocks located in odd (2-P.B.O) and even (2-P.B.E) links with the rest of system.} \label{EE-Na-3}
\end{figure}

As previously mentioned the 2nd region of the GSPD has twice as many critical fields as the 3rd and 4th regions do. For this area of GSPD the EE between a single-particle and the rest of the system is shown in the Fig.~\ref{EE-Na-1}(a), as well as, the treatment of EE related to the three-particle block is available too. In the absence of a magnetic field; it is clearly seen that a single-particle, independent on its location in the chain, is entangled with ROC. With increasing of a magnetic field, up to the first critical field, this situation will remain. At the first critical field, a severe reduction of the entanglement begins. This decrement is continuous until the magnetic field reaches to the 2nd critical point. This point has less reduction rate than the first critical point and for the bigger values of magnetic fields, the entanglement will be vanished.
\begin{figure}[t]
\centerline{\psfig{file=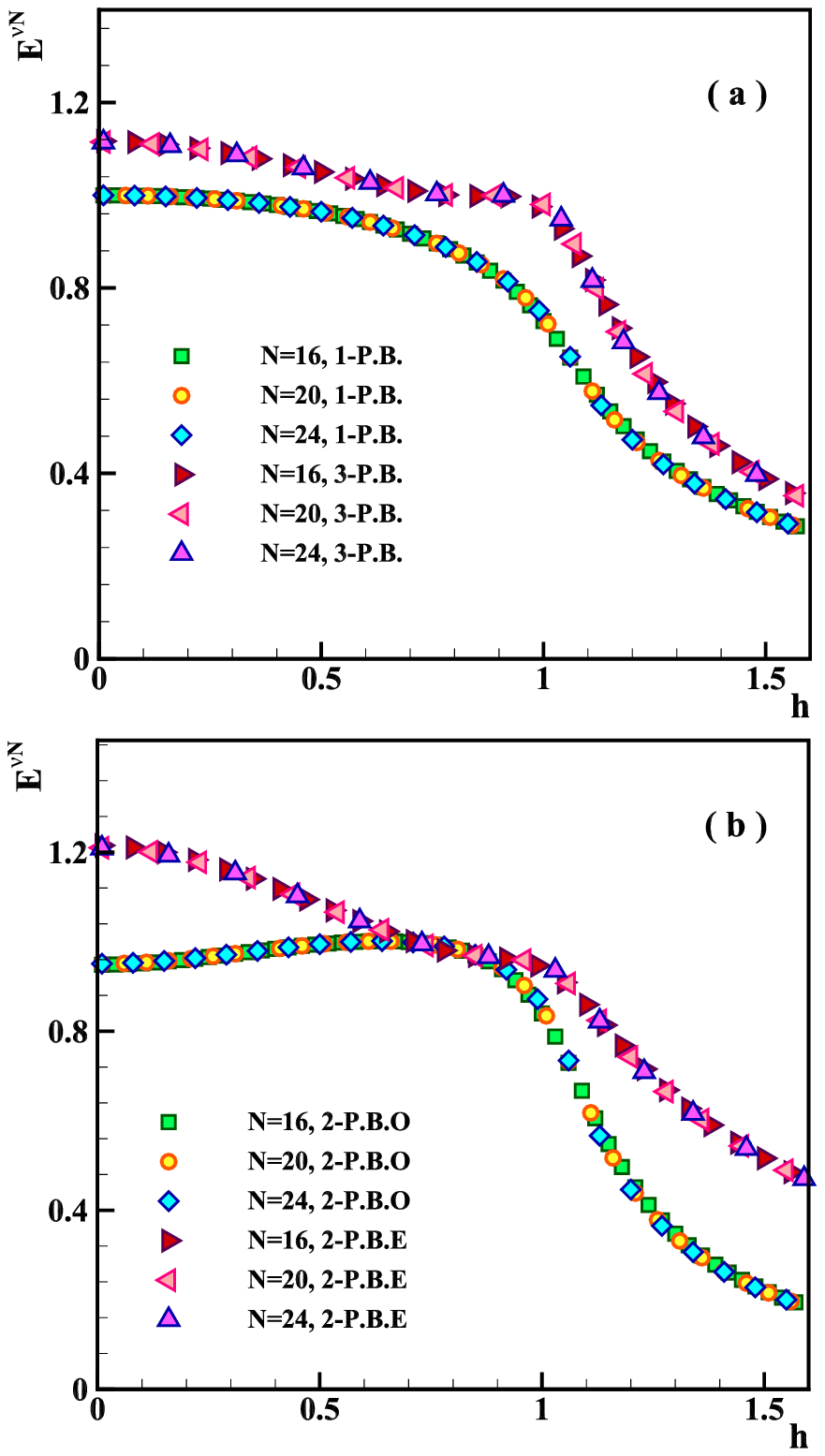,width=8.0cm,height=15.0cm}}
\caption{(Color online.)Entanglement of entropy as a function of the transverse field $h$ for different chain lengths
    $N=16,20,24$ and exchanges $L_{1}=1.0$, $J_1=3.0, J_2=0.5$ (a) entanglement of one-particle block (1-P.B.) and three-particle block (3-P.B.)with the rest of system. (b) entanglement of two-particle blocks located in odd (2-P.B.O) and even (2-P.B.E) links with the rest of system.} \label{EE-Na-2}
\end{figure}
Also a three-particle block is entangled with ROC in the fields less than $h_{c_{1}}\approx1.2$, both parts of the system tend to be more entangled for $h_{c_{1}}< h< h_{c_{2}}$. However, for $h_{c_{2}}\approx1.7<h$, EE will be over.

Another main point is the different behavior of the entanglement of entropy for various two-particle blocks with the ROC. As shown in the Fig.~\ref{EE-Na-1}(b), this behavior depends on where a two-particle block is. In the absence of a field, if a block is located at an odd link with couplings ($J_{1},J_{2}$), it almost will not be entangled with the ROC, otherwise a two-particle block in an even link, with coupling ($L_{1}$), will be entangled with ROC.

Also, amount of entanglement of a two-particle block in an odd link is the same as that of an even link in a factorized point\cite{motamedi12} independent on where the blocks are. This point approximately occurs at $h_{f}\approx 1.7$.

In the sequence of surveying the EE, we will receive to the 3rd and 4th regions of GSPD. One- and three-particle blocks entanglement have qualitatively similar behavior for the 3rd and 4th regions. As presented in Fig.~\ref{EE-Na-3}(a) and Fig.~\ref{EE-Na-2}(a) one-particle block entanglements for these regions are recognizable to the three-block entanglement since for both of regions one-particle block starts from a value near $1$ and reduces from their critical fields to vanish in the larger magnetic field.

However, three-particle block for both of 3rd and 4th regions has the larger entanglement amount than the one-particle block. In addition, cusps near critical fields as presented in Fig.~\ref{EE-Na-3}(a) and Fig.~\ref{EE-Na-2}(a) specify the three-particle block entanglement.

Beside, for the 3rd and 4th regions odd- and even-block entanglements have almost near value (Fig.~\ref{EE-Na-3}(b) and Fig.~\ref{EE-Na-2}(b)) , in contrast with the 1st and 2nd regions in which the difference of the values even reaches to hundred times. It is because, the value of $J_{2}$ in odd-links for the 3rd and 4th regions is much less than that of the 1st and 2nd regions. Furthermore, in the 4th area of GSPD we can observe a factorized point (like 2nd area) in which amount of entanglement for even and odd-links are the same\cite{motamedi12}. For this area, the factorized point is approximately equals to $h_{f}\approx0.8$.

\section{conclusion}\label{sec-III }
In this paper, we have considered the 1D quantum compass model with the periodic boundary conditions in the presence of an external transverse magnetic field. By using the exact diagonalization approach, we obtained some magnetic response functions at zero temperature. First, we have calculated the fidelity and fidelity susceptibility. Our computation of the fidelity shows that this quantity has maximum value ($F\approx 1$) for all fields, and some abrupt drops are observed as the magnetic field reaches near a quantum critical point. Dropping in fidelities raises the cusps in fidelity susceptibilities. We show that the difference between the position of the sharp drop of fidelity $h^{*}$ from real critical field $h_{c}$ is inversely related to the number of particles with a relation such as $h^{*}=h_{c}+\Sigma_{j} c_{j} N^{-1/\nu}$ for a finite chain. In this relation $\nu$ is the correlation length critical exponent. The scaling behavior of the extremum of fidelity susceptibility shows that the amount of $\nu$ depends on the selected area of GSPD. From the scaling relation governing to pseudo-critical points, a linear dependence for the maximum fidelity susceptibilities $\chi^{max}_{F}$ on $N^{2/d\nu}$ is obtained. The critical exponents $\nu$ obtained from the finite size scaling are different for the various regions of the GSPD of QCC. As a matter of fact, because of the different employed method to obtain these results, they are a bit different from those of Ref.[25]. Also it was observed that fidelity susceptibility per site or averaged fidelity susceptibility ($\chi_{F}/N$) is an intensive quantity in the off-critical fields and extensive in the critical fields.

On the other hand, we have looked at the Von-Neumann entropy for one-, two- or three-particle blocks in the chain and presented that a two-particle block located at an odd link can be entangled with the ROC, hundred times as much as a two-particle block of an even link. It indicates the value of entanglement between a two-particle block with the rest of the system is more dependent on the power of exchange couplings connecting the block with the rest of the system than the power of exchange coupling between two particles in the own block. In facts, a pair with a strong coupling does not see the rest of the system. Also, in the 2nd and 4th region of GSPD, amount of entanglement of a two-particle block in an odd link is the same as that of an even link in a factorized point independent on where the block is.

\section{acknowledgments}
We are grateful to Faezeh Mohammadbeigi from Simon Fraser University for her fruitful, helpful and friendly aid for preparing resources.

\vspace{0.3cm}


\begin{thebibliography}{99}



\bibitem{cheong07}
S.W. Cheong, Nature Mater. 6, 927 (2007).
\bibitem{tokura00}
Y. Tokura and N. Nagaosa, Science 288, 462 (2000).
\bibitem{wakabayashi06}
Y. Wakabayashi, D. Bizen, H. Nakao, Y. Murakami, M. Nakamura, Y. Ogimoto, K. Miyano, and H. Sawa, Phys. Rev. Lett. 96, 017202 (2006).
\bibitem{Wen-Long2011}
Wen-Long You and You-Li Dong, Phys. Rev. B. 84, 174426 (2011).
\bibitem{mishra2004}
Anup Mishra, Michael Ma, Fu-Chun Zhang, Siegfried Guertler, Lei-Han Tang, and Shaolong Wan,  Phys. Rev. Lett. 93, 207201 (2004).
\bibitem{kugel73}
K. I. Kugel and D. I. Khomskii, Sov. Phys. Jetp 37, 725 (1973).
\bibitem{jackeli2009}
G. Jackeli and G. Khaliullin, Phys. Rev. Lett. 102, 017205 (2009).
\bibitem{kim2008}
B.J. Kim, H. Jin, S.J. Moon, J.-Y. Kim, B.-G. Park, C.S.leem, J. Yu, T.W. Noh, C. Kim, S.-J. Oh, J.-H. Park, V.Durairaj, G. Cao, E. Rotenberg, Phys.Rev. Lett. 101,076402 (2008)
\bibitem{khomski2003}
Khomskii D.I. and Mostovoy M. V., J. Phys. A. 36, 9197, (2003).
\bibitem{nussinov05}
Z. Nussinov and E. Fradkin, Phys. Rev. B 71, 195120 (2005).
\bibitem{katsura2005}
Hosho Katsura, Naoto Nagaosa, and Alexander V. Balatsky, Phys. Rev. Lett. 95, 057205 (2005).
\bibitem{murakami2004}
Shuichi Murakami, Naoto Nagaosa, and Shou-Cheng Zhang, Phys. Rev. Lett. 93, 156804 (2004).
\bibitem{Breziskii2007}
Wojciech Brzezicki, Jacek Dziarmaga, Andrzej M. Ole\'{s}, Rev. B. 75, 134415 (2007).
\bibitem{erik09}
Erik Eriksson, and Henrik Johannesson, Phys. Rev. B 79, 224424 (2009).
\bibitem{oles06}
A. M. Ole\'{s}, P. Horsch, L. F. Feiner, and G. Khaliullin, Phys. Rev. Lett. 96, 147205 (2006).
\bibitem{mahdavifar10}
S. Mahdavifar, Eur. Phys. J. B 77, 77-82 (2010).
\bibitem{sachdev11}
S. Sachdev, Quantum Phase Transitions (Cambridge University Press, Cambridge, U.K., 2011).
\bibitem{osterloh02}
A. Osterloh et al., Nature 416, 608 (2002); G. Vidal et al., Phys. Rev. Lett. 90, 227902 (2003); L. Amico et al., Rev.Mod. Phys. 80, 517 (2008).
\bibitem{vidal03}
G. Vidal et al., Phys. Rev. Lett. 90, 227902 (2003); J.I. Latorre, E. Rico and G. Vidal, Quant. Inf. Comput. 4,
48 (2004).
\bibitem{bennett96}
C.H. Bennett et al., Phys. Rev. A 53, 2046 (1996).
\bibitem{wootters81}
W. K. Wootters, Phys. Rev. D \textbf{23}, 357 (1981).
\bibitem{josa94}
R. Josa, J. Mod. Opt. 41, 2315 (1994).
\bibitem{schumacher95}
B. Schumacher, Phys. Rev. A 51, 2738 (1995).
\bibitem{ke-wei2009}
Ke-Wei Sun and Qing-Hu Chen, Phys. Rev. B 80, 174417 (2009).
\bibitem{jafari11}
R.Jafari, Phys. Rev. B 84, 035112 (2011).
\bibitem{motamedi11}
M. Motamedifar, S. Mahdavifar, S. F. Shayesteh, Eur. Phys. J. B 83, 181-189 (2011).
\bibitem{w-L2007}
Wen-Long. You, Y.-W. Li, and S.-J. Gu, Phys. Rev. E 76, 022101 (2007).
\bibitem{quan06}
H. T. Quan, Z. Song, X. F. Liu, P. Zanardi, and C. P. Sun, Phys. Rev. Lett. 96, 140604 (2006).
\bibitem{zanardi06}
P. Zanardi and N. Paunkovic, Phys. Rev. E 74, 031123(2006).
\bibitem{gu2010}
S. Gu. International Journal of Modern Physics B 24, 4371 (2010)
\bibitem{zanardi2007}
Lorenzo Campos Venuti, and Paolo Zanardi, Phys. Rev. Lett 99, 095701(2007)
\bibitem{capponi2009}
David Schwandt, Fabien Alet, and Sylvain Capponi, Phys. Rev. Lett 103, 170501 (2009)

\bibitem{capponi2010}
A. Fabricio Albuquerque, Fabien Alet, Clément Sire, and Sylvain Capponi, Phys. Rev. B 81, 064418 (2010)

\bibitem{zanardiA2008}
Paolo Zanardi, Matteo G. A. Paris and L. Compos Venuti, Phys. Rev. A 78, 042105 (2008).
\bibitem{sandvik2011}
C. De Grandi, A. Polkovnikov, A. W. Sandvik, Phys. Rev. B 84, 224303 (2011)
\bibitem{motamedi12}
M. Motamedifar, S. Mahdavifar, S. Farjami Shayesteh, being prepared to be published.







\end{thebibliography}
\end{document}